\documentclass[printer]{tJOT2e}
\newcommand{\Pm}{\mathrm{Pm}}
\newcommand{\Ru}{\mathrm{Re}}
\newcommand{\Rm}{\mathrm{Rm}}

\usepackage{rotating}
\begin{document}
\issnp{} \jvol{00} \jnum{00} \jyear{2006} \jmonth{January}

\markboth{Rodion Stepanov and Franck Plunian}{Dynamo action at low Pm}

\title{\itshape Fully developed turbulent dynamo at low magnetic Prandtl numbers.}

\author{
Rodion Stepanov $\dagger$ \\ 
Institute of Continuous Media Mechanics,\\
 Korolyov 1, 614013 Perm, Russia\\
Franck Plunian $\ddagger$ \\ 
Laboratoires des Ecoulements G\'{e}ophysiques et
Industriels,\\
B.P. 53, 38041 Grenoble Cedex 9, France
\thanks{
\vspace{6pt}
\newline{\tiny{ {\em } $\dagger$ rodion@icmm.ru}}
\newline{\tiny{ $\ddagger$ Franck.Plunian@hmg.inpg.fr; \quad http://legi.hmg.inpg.fr/$\sim$
plunian}}
}}  
\received{\today}

\maketitle

\begin{abstract}
We investigate the dynamo problem in the limit of small magnetic Prandtl number ($\Pm$) using a shell model of magnetohydrodynamic turbulence. The model is designed to satisfy conservation laws of total energy, cross helicity and magnetic helicity in the limit of inviscid fluid and null magnetic diffusivity. The forcing is chosen to have a constant injection rate of energy and no injection of kinetic helicity nor cross helicity. We find that the value of the critical magnetic Reynolds number ($\Rm$) saturates in the limit of small $\Pm$. Above the dynamo threshold we study the saturated regime versus $\Rm$ and $\Pm$. In the case of equipartition, we find Kolmogorov spectra for both kinetic and magnetic energy except for wave numbers just below the resistive scale. Finally the ratio of both dissipation scales (viscous to resistive) evolves as $\Pm^{-3/4}$ for $\Pm < 1$.
\end{abstract}
\section{Introduction} \label{intro}
Most of astrophysical bodies possess or have had in their history
their own magnetic fields. In most cases their generation rely on
inductive processes produced by the turbulent motion of the
electroconducting fluid within the body \cite{Rudiger04}. An
important parameter of the problem is the magnetic Prandtl number
defined by $\Pm=\nu / \eta$ where $\nu$ is the viscosity and
$\eta$ the magnetic diffusivity of the fluid. In the ``magnetic"
universe $\Pm$ varies from values as large as $10^{14}$ for the
interstellar medium \cite{Schekochihin04b} to values as small as $10^{-6}$ for the iron
core of planets or stellar plasmas. This large spectrum of
possible $\Pm$ values implies strong differences between possible
generation mechanisms. In some sense $\Pm$ is a measure of the
kinetic energy spectrum available for generating magnetic energy.
When $\Pm \ge 1$ the resistive scale is smaller than the viscous
scale implying that all velocity scales are available for
generating some magnetic field. In the other hand for $\Pm < 1$,
only the velocity scales larger than the resistive scale are
available for the magnetic field generation. In that case, the
velocity scales smaller than the resistive scale are enslaved to
the larger scales and in essence they stay passive in the
generation process. Besides this is why the large eddy simulation
technique may be recommended in that case \cite{Ponty05}. Therefore,
at first sight one can expect that dynamo action is all the more
difficult to obtain since $\Pm$ is smaller in reason of a smaller
velocity spectrum available for the magnetic generation. 
This is
indeed what comes out from recent numerical simulations
\cite{Nordlund92,Brandenburg96b,Nore97,Christensen99,Yousef03,Scheko04}
(see also \cite{Rogachevskii04} and references therein for an alternative approach). Though, we
have evidence of magnetic field in planets and stars, and dynamo
action has also been reproduced in experiments working with liquid
sodium for which $\Pm$ is small ($\sim 10^{-6}$)
\cite{Gai00,Gai01,Sti01,Muller04}. These
experiments and further devices in preparation \cite{Bourgoin02,Ravelet05,Frick02}
are designed in such a way that the dynamo mechanism is produced
by the large scale of the flow due to an appropriate large scale
forcing. The turbulence naturally developing at smaller scales may
play a role though this is still unclear \cite{Normand03,Leprovost04,Ponty05,Stepanov05,Dubrulle06}. In these
experiments, the choice of the forcing is based on the hypothesis
that it is the stationary part of the large scale flow which
should be important for the generation mechanism. A number of flow
geometries studied in the past turned out to be good candidates
for such experiments \cite{Roberts72,Ponomarenko73,Dudley89}.

In the present paper we are interested in the possibility for a
Kolmogorov type turbulent flow to generate dynamo action at low
$\Pm$, without need for a large scale motion controlling the
generation mechanism. We expect the eddies having the highest
shearing rate to be the more active for generating the magnetic
field, at least during the kinematic stage of magnetic field
growth. As in Kolmogorov turbulence $u_l/l \approx l^{-2/3}$,
these eddies correspond to the smallest available scale which is
the viscous scale for $\Pm \ge 1$ \cite{Kulsrud92} and the
resistive scale for $\Pm < 1$ \cite{Scheko04}. Eventually the
magnetic field will then spread out to larger scales due to the
nonlinear interactions. This problem is hard to solve by direct
numerical simulation for it needs high resolution in order to
describe magnetic phenomena adequately \cite{Boldyrev04}. Some
results have been obtained using the EDQNM closure applied to the
MHD equations \cite{Leorat81} near the critical $\Rm$ and for
arbitrary low values of $\Pm$. Here we want to investigate
arbitrary large values of $\Rm$ and small values of $\Pm$. For
that we use a shell model of MHD turbulence introduced by Frick
and Sokoloff \cite{Frick98}. This model is the successor of
several other shell models for MHD turbulence \cite{Frick83,Gloaguen85,Grappin86,Carbone90,Carbone94,Biskamp94,Brandenburg96a}  but it is
the only one to conserve all integrals of motions including
magnetic helicity (or kinetic helicity for the non magnetic case).
It is based on the so-called GOY hydrodynamic shell model
\cite{Gledzer73,Yamada87,Frisch95,Biferale03}. In \cite{Frick98}, Frick and
Sokoloff have derived a model which represents either 2D or 3D MHD
turbulence, depending on the choice of two parameters. As in real
MHD turbulence the 2D model leads to the impossibility of dynamo
action \cite{Zeldovich56}. This shows that in spite that such a
shell model is a drastic simplification of the real MHD
turbulence, ignoring for example the geometrical structures of the
motion and magnetic field, it contains enough features
to make the difference between the 2D and 3D problems (see also \cite{Giulani98}). It also
reproduces quite well the structure functions at different orders
of real MHD turbulence. Here we consider only the 3D model
herein after referred to as FS98. This model has also been used by Lozhkin \textit{et al.} \cite{Lozhkin99}
to show that small scale dynamo is possible at low $\Pm$,
contrary to the hypothesis put forward by Batchelor \cite{Batchelor50}.

Giulani and Carbone \cite{Giulani98}
have shown that long runs with the FS98 model
lead inevitably towards a ``dynamical alignement'' stopping the nonlinear transfer
towards the smaller scales. Giulani and Carbone \cite{Giulani98}
suggested that this problem might
be overcome with an other choice of the external driving force.
This is what we have done here, adopting a forcing in such a way that
it acts on several scales and depends on time with a random phase
at each forcing scale (see section \ref{sec:forcing}).
Finally,
we took care to have long runs well beyond any transient
state, in order to have good statistics and reliable results.
\section{Shell model for MHD turbulence}\label{model}
\subsection{Model equations}
The shell model is built up by truncation of the Navier-Stokes and
induction equations. We define logarithmic
shells, each shell being characterized by one real wave number
$k_n=k_0\lambda^{n}$ and dynamical complex quantities $U_n$ and
$B_n$ representative of the velocity and magnetic fluctuations for
wave vectors of norm ranging between $k_n$ and $k_{n+1}$. The
parameter $\lambda$ is taken equal to the gold number
$(1+\sqrt{5})/2$ for it optimizes the resolution \cite{Sabra98}.
The model is described by the following set of equations ($0 \le n
\le N$)
\begin{eqnarray}
d_t U_n &=& i k_n (Q_n(U,U,a)-Q_n(B,B,a))\nonumber\\
 &-& \nu k_n^2 U_n + F_n, \label{goy_u} \\
d_t B_n &=& i k_n (Q_n(U,B,b)-Q_n(B,U,b)) \nonumber\\
&-& \eta k_n^2 B_n , \label{goy_b}
\end{eqnarray}
where
\begin{eqnarray}
Q_n(X,Y,c)&=& c_1 X_{n+1}^*Y_{n+2}^*  + c_2 X_{n-1}^*Y_{n+1}^*\nonumber \\
&+& c_3 X_{n-2}^*Y_{n-1}^*.
\end{eqnarray}
represents the nonlinear transfer rates with the four neighbouring
shells $n-2$, $n-1$, $n+1$ and $n+2$. In addition we have to take
$U_{-2}=U_{-1}=U_{N+1}=U_{N+2}=0$ and
$B_{-2}=B_{-1}=B_{N+1}=B_{N+2}=0$. The parameter $F_n$ is the
forcing at shell $n$. The time unit is defined by the turnover
time of the largest scale $\tau=(|U_0| k_0)^{-1}$. To determine the
complex coefficients $a_j$ and $b_j$, $j=1,2,3$ we apply the
property that the total energy $E_{tot}$, cross-helicity ${\cal H}_C$
and magnetic helicity ${\cal H}_B$ must be conserved in the limit
of non-viscous and non-resistive limit $\nu = \eta = 0$. In our
shell model, these quadratic quantities write in the following
form
\begin{eqnarray}
    E_{tot}&=&\frac{1}{2}\sum_{n=0}^N (|U_n|^2 + |B_n|^2), \label{totalE}\\
    {\cal H}_C&=&\frac{1}{2}\sum_{n=0}^N (U_n B_n^*+B_n U_n^*),\label{crosshelicity}\\
     {\cal H}_B&=&\frac{1}{2}\sum_{n=0}^N (-1)^n |B_n|^2/k_n,\label{maghelicity}
\end{eqnarray}
leading to $a_1=1$, $a_2=(1-\lambda)\lambda^{-2}$,
$a_3=-\lambda^{-3}$, $b_1=b_2=b_3=(\lambda(1+\lambda))^{-1}$. In
the pure hydrodynamic case ($B_n=0$) the original GOY model is
recovered satisfying, in addition to (\ref{totalE}), the
conservation of the kinetic helicity \cite{Kadanoff95}
\begin{equation}
{\cal H}_U=\frac{1}{2}\sum_{n=0}^N (-1)^n |U_n|^2 k_n.
\label{kinhelicity}
\end{equation}
\subsection{Forcing and initial conditions}
\label{sec:forcing}
The forcing is chosen in order to control the injection rate of
kinetic energy, cross and kinetic helicities. For that we spread
the forcing on three neighbouring shells $n_f, n_f+1$ and $n_f+2$
with $F_{n_f+j} = f_j e^{i\phi_j}$, $j=0,1,2$ where the $f_j$ are
positive real quantities and where the $\phi_j \in \left[ 0, 2\pi
\right]$ are random phases. In that case the forcing is
$\delta$-correlated. Alternatively we also used a forcing for
which the phases $\phi_j$ are constant during a certain time
$\tau_c$, which can be interpreted as a finite correlation time.
In fact this does not make much difference either on the
autocorrelation functions of $U_n$ nor on the
subsequent results. Therefore it is sufficient to use random
phases. As we are interested to inject neither kinetic helicity
nor cross-helicity, the forcing functions must satisfy
\begin{eqnarray}
\label{forcing}
  \frac{1}{2}\sum_{n=n_f}^{n_f+2} U_{n}^* F_n+ U_{n} F_n^* = \varepsilon, \label{forcing1}\\
  \sum_{n=n_f}^{n_f+2} (-1)^n k_{n} (U_{n}^* F_n+ U_{n} F_n^* ) = 0, \label{forcing2}\\
  \sum_{n=n_f}^{n_f+2} B_{n}^* F_n + B_{n} F_n^*  =  0. \label{forcing3}
\end{eqnarray}
where $\varepsilon$ is the rate of kinetic energy supplied to the
system. Therefore for a given set of random $\phi_j$ ($j=0,1,2$),
the $f_j$  depend on the $U_j$ and $B_j$ ($j=0,1,2$) which
expressions are given in Appendix. For some arbitrary initial
conditions on $U_j$ ($j=0,1$) of small intensity ($\sim 10^{-6}$)
we let the hydrodynamic evolve until it reaches some statistically
stationary state. Then introducing at a given time some arbitrary
non zero values of $B_j$ ($j=0,1$) of small intensity ($\sim
10^{-6}$) we solve the full problem until a statistically
stationary MHD state is reached. The time of integration needed to
obtain good statistics depends on $\nu$ and $\eta$ but typically it is equal to
several hundreds of the large scale turn-over time.
\subsection{Input and output}
\label{Inandout}
The input parameters of the problem are $\nu$, $\eta$, the forcing
shell $n_f$, the rate of injected kinetic energy $\varepsilon$ and
the number of shells $N$. In the rest of the paper we take
$\varepsilon=1$. 

As output we define the kinetic and magnetic
energy for the shell $n$ by
\begin{equation}
\label{local} E^U(n)=\frac{1}{2}|U_n|^2 \quad \mbox{and} \quad
E^B(n)=\frac{1}{2}|B_n|^2,
\end{equation}
the total kinetic and total magnetic energy by
\begin{equation}
\label{global} E_U=\sum_{n=0}^N E^U(n) \quad \mbox{and} \quad
E_B=\sum_{n=0}^N E^B(n)
\end{equation}
and the total energy by
\begin{equation}
E_{tot} = E_U + E_B.
\end{equation}
 Following \cite{Verma04} we define the spectral energy fluxes from
the inside of the $U$(or $B$)-sphere (shells with $k<k_n$) to the
outside of the $U$(or $B$)-sphere (shells with $k \ge k_n$). We
note for example $\Pi_{U>}^{B<}(n)$ the energy flux from the
inside of the $B$-sphere to the outside of the $U$-sphere. Then we
have
\begin{eqnarray}
    \Pi_{U>}^{U<}(n)&=&\sum^{n-1}_{j=0}\Im\{k_j U_j^*Q_j(U,U,a)\}\\
    \Pi_{U>}^{B<}(n)&=&\sum^{n-1}_{j=0}\Im\{-k_j U_j^*Q_j(B,B,a)\}\\
    \Pi_{B>}^{U<}(n)&=&\sum^{n-1}_{j=0}\Im\{-k_j B_j^*Q_j(B,U,b)\}\\
    \Pi_{B>}^{B<}(n)&=&\sum^{n-1}_{j=0}\Im\{k_j B_j^*Q_j(U,B,b)\}.
\end{eqnarray}
In FS98 the time average of $\Pi_{U>}^{U<}(n)$ is denoted $\Pi_n$.
We also define the energy fluxes from the inside of the $U$-and-$B$-spheres to the outside of the $U$-sphere or $B$-sphere
by
\begin{eqnarray}
    \Pi_{U}(n)&=&\Pi_{U>}^{U<}(n)+\Pi_{U>}^{B<}(n)\\
    \Pi_{B}(n)&=&\Pi_{B>}^{U<}(n)+\Pi_{B>}^{B<}(n)
\end{eqnarray}
and the total energy flux by
\begin{eqnarray}
    \Pi_{tot}(n)&=&\Pi_{U}(n)+\Pi_{B}(n).
\end{eqnarray}
We define the viscous and resistive dissipation rates $D^U(n)$ and
$D^B(n)$ in shell $n$, by
\begin{eqnarray}
\label{dissipation}
  D^U(n)&=&  \nu k_n^2 |U_n|^2\\
  D^B(n)&=&  \eta k_n^2 |B_n|^2
\end{eqnarray}
and the total dissipation rate by
\begin{eqnarray}
D_{tot}=\sum_{n=0}^N (D^U(n)+D^B(n)).
\end{eqnarray}
With these definitions we obtain the following shell-by-shell
energy budget equations:
\begin{eqnarray}
d_t\sum_{j=0}^n E^U(j)+\Pi_{U}(n) &=&
 -\sum_{j=0}^n D^U(j)+\epsilon \label{budgetU}\\
d_t\sum_{j=0}^n E^B(j)+\Pi_{B}(n) &=& -\sum_{j=0}^n D^B(j).\label{budgetB}
\end{eqnarray}

For a statistical stationary solution ($d_t\langle E^U(j)\rangle
=d_t\langle E^B(j)\rangle =0$) we have then
\begin{eqnarray}
\label{pitot} \langle \Pi_{tot}(n)\rangle &=&
 -\sum_{j=0}^n \langle D^U(j)\rangle -\sum_{j=0}^n \langle D^B(j)\rangle
+\epsilon.
\end{eqnarray}
where here and after $\langle \quad \rangle $ denotes time
averaged quantities.

We define the kinetic and magnetic Reynolds numbers as
\begin{eqnarray}
\label{reynolds}
  \Ru &=&  \langle E_{tot}\rangle ^2/(\nu \langle D_{tot}\rangle )\\
  \Rm &=&  \langle E_{tot}\rangle ^2/(\eta \langle D_{tot}\rangle ).
\end{eqnarray}

Finally, following \cite{Kraichnan67}, we
define the viscous (resp. resistive) scale $k_{\nu}^{-1}$
(resp. $k_{\eta}^{-1}$) as the one at which the viscous
(resp. Ohmic) decay time $\tau_{\nu}=(\nu k_n^2)^{-1}$ (resp. $\tau_{\eta}=(\eta k_n^2)^{-1}$) becomes comparable to
the typical turn-over time $\tau_U=(k_n \langle |U_n|^2 \rangle^{1/2})^{-1}$.

\section{Hydrodynamics}\label{hyd}
\begin{figure}
  \begin{tabular}{@{}c@{\hspace{0em}}c@{\hspace{0cm}}l@{}}
    \includegraphics[width=0.46\textwidth]{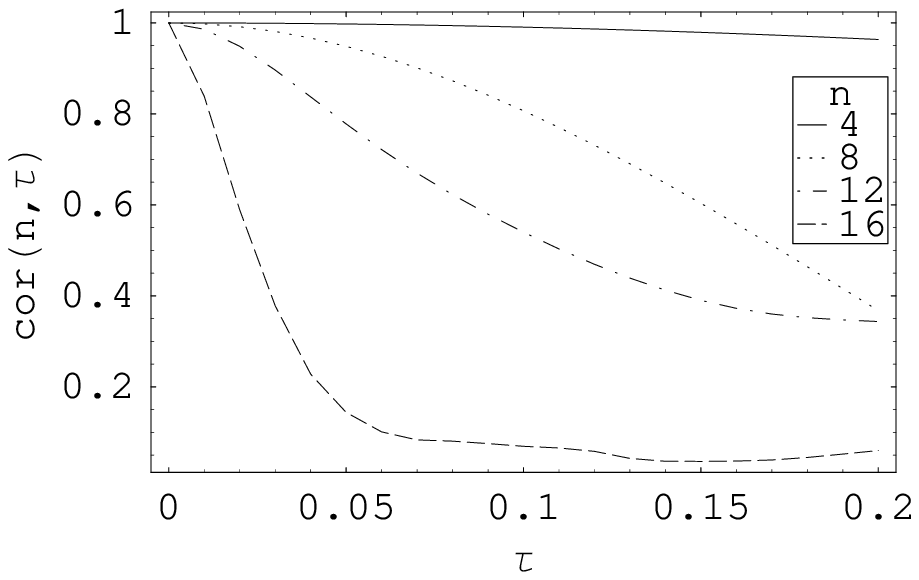}
    &
    \includegraphics[width=0.44\textwidth]{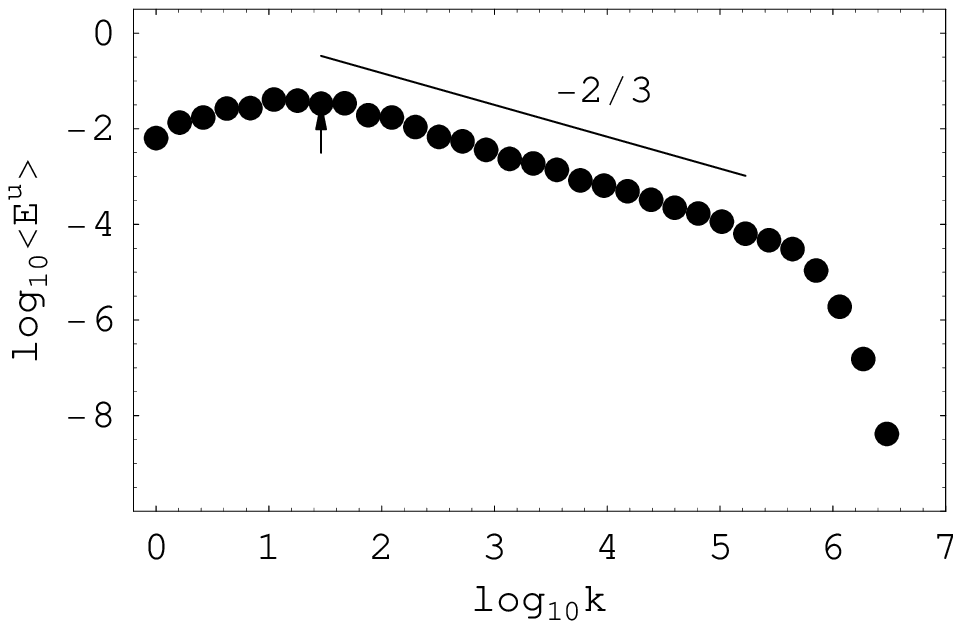}
    &
    \\*[-2.cm]
    (a)&(b)&\\*[2.cm]
    \includegraphics[width=0.45\textwidth]{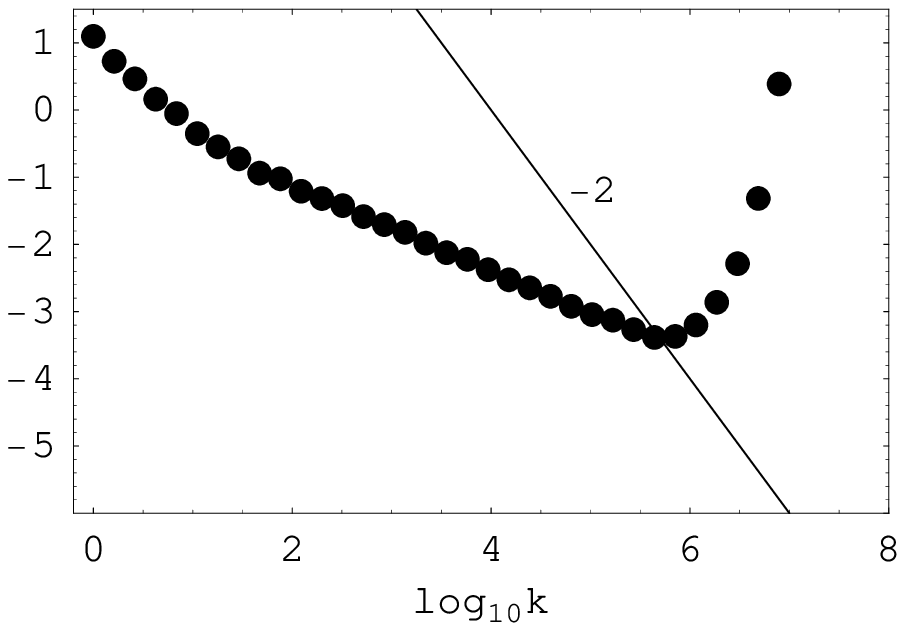}
    &
    \includegraphics[width=0.44\textwidth]{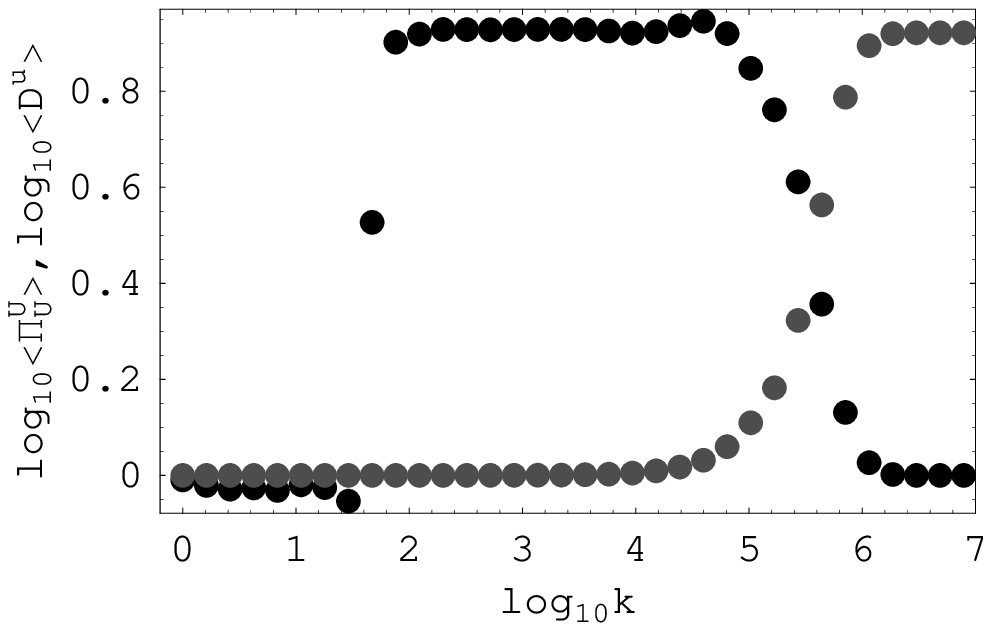}
    &
        \\*[-2.cm]
    (c)&(d)&\\*[1.5cm]
  \end{tabular}
\caption{Hydrodynamic case for $\nu=10^{-8}$ and a forcing scale (arrow) corresponding
to $n_f=8$. The output Reynolds number is $\Ru=8 \;10^7$. In (a), the autocorrelation function $cor(n,\tau)$ for a $\delta$-correlated
forcing is plotted versus $\tau$ and for several shells $n$.
In (c), the turn-over (black dots) and dissipation (straight line) characteristic times are plotted
versus $\log_{10} k$. In (b), the energy spectrum is plotted versus $\log_{10} k$ and the $k^{-2/3}$ slope (full line) is plotted for comparison. In (d), the energy flux (black dots) and
the dissipation $\sum_{j=0}^n D^U(j)$ (gray dots) are plotted versus $\log_{10} k$.}
\label{fig1}
\end{figure}

Choosing the appropriate forcing corresponding to $B_n=0$ we
present in Fig.\ref{fig1} some results concerning the pure
hydrodynamic case for $\nu=10^{-8}$ and $n_f=8$.
In this case the forcing is $\delta$-correlated. Though the
autocorrelation function, defined by
\begin{equation}\label{ecor}
    cor(n,\tau)=\frac{\int U_n^*(t)U_n(t+\tau)+U_n(t)U_n^*(t+\tau)
    dt}{2\sqrt{\int U_n^*(t)U_n(t) dt \int U_n(t+\tau)U_n^*(t+\tau) dt}}
\end{equation}
and plotted in Fig.\ref{fig1}a, is far from being
the one of a $\delta$-correlated velocity contrary to the Kasantzev model \cite{Scheko04}.
We also made comparisons with a finite correlation time
forcing without finding any significant differences. Therefore the
$\delta$-correlated forcing does not seem to be an issue in our
problem. 

The kinetic
energy spectrum (Fig.\ref{fig1}b, black dots) of the stationary statistical state is found to be in
$k^{-2/3}$ (which corresponds to a Fourier energy spectrum
of $k^{-5/3}$ as expected in Kolmogorov turbulence).
In Fig.\ref{fig1}c, the spectral flux $\Pi_U(n)$ (black dots)
and the dissipation $\sum_{j=0}^n D^U(j)$ (gray dots) are found to satisfy the kinetic energy budget (\ref{budgetU}) with $\epsilon=1$.
In addition, in the inertial range we find that $\Pi_U(n) \sim \epsilon$ and $\sum_{j=0}^n D^U(j) \sim 0$ as predicted by a Kolmogorov turbulence. After the viscous scale,
$\Pi_U(n) \sim 0$ and $\sum_{j=0}^n D^U(j) \sim \epsilon$.

As previously defined, the viscous scale 
is the one at which the viscous
decay time $\tau_{\nu}=(\nu k_n^2)^{-1}$ (full curve of Fig.\ref{fig1}c) becomes comparable to
the typical turn-over time $\tau_U=(k_n U_n)^{-1}$ (black dots of Fig.\ref{fig1}c). This leads to $k_{\nu} \sim 10^6$ and compares
indeed very well with the Kolmogorov dissipation scale 
$k_{\nu}^{-1}\sim (\nu^3 / \epsilon)^{1/4}$.
Finally the little bump of $\Pi_U(n)$  (black dots Fig.\ref{fig1}d) just before the viscous scale looks like a bottle-neck effect \cite{Dobler03}.
\section{Dynamo action}\label{dyn}
\subsection{Time evolution of quadratic quantities}
\begin{figure}
  \begin{tabular}{@{}c@{\hspace{0em}}c@{\hspace{0em}}c@{}}
  \includegraphics[width=0.33\textwidth]{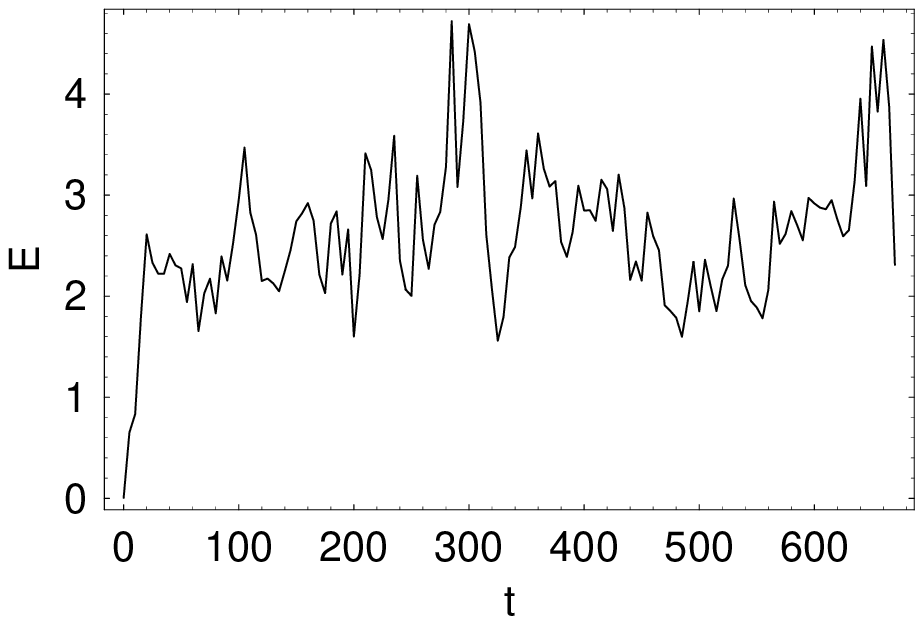}
    &
    \includegraphics[width=0.33\textwidth]{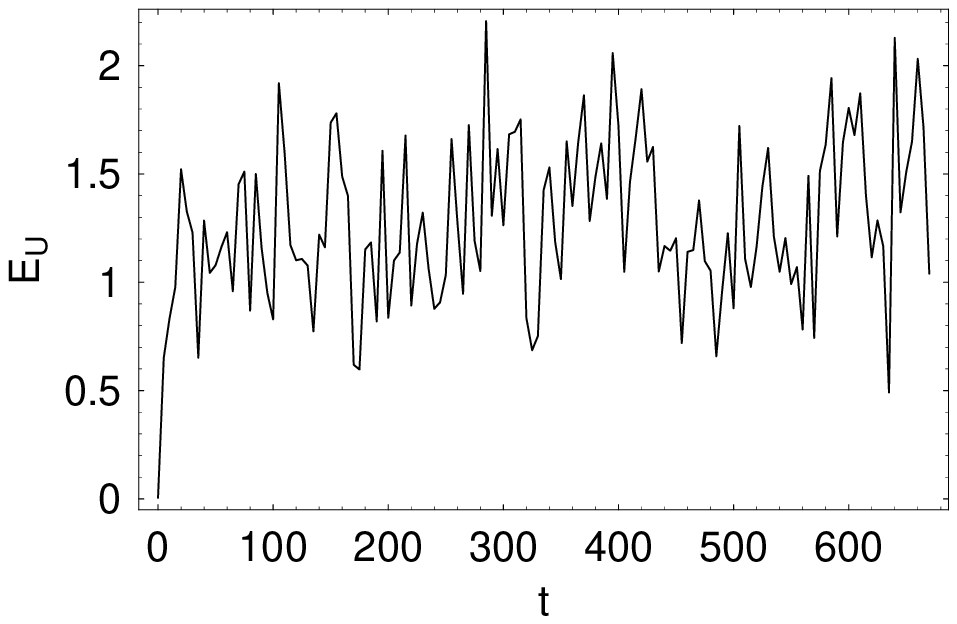}
    &
    \includegraphics[width=0.33\textwidth]{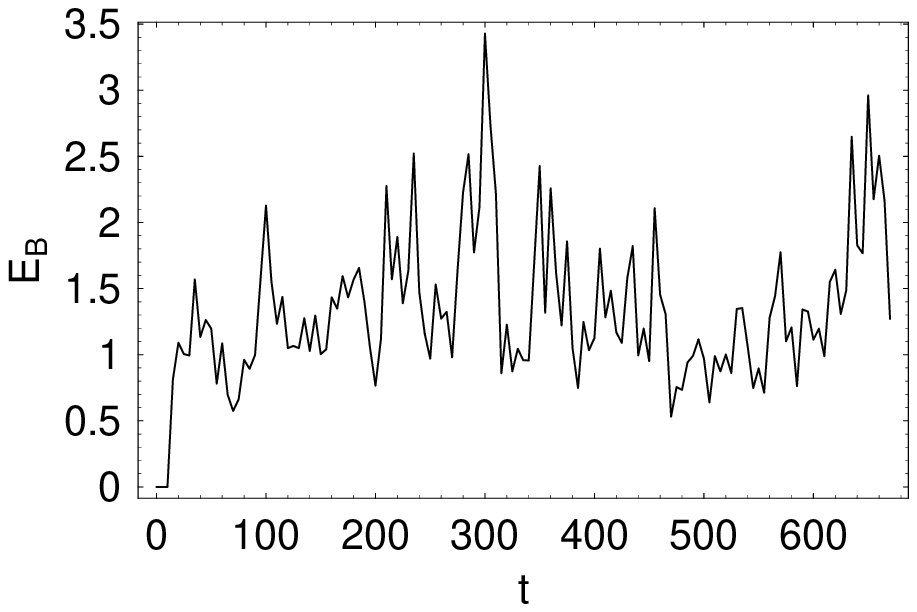}
    \\*[-0.3cm]
(a)&(b)&(c)\\*[0.3cm]
    \includegraphics[width=0.33\textwidth]{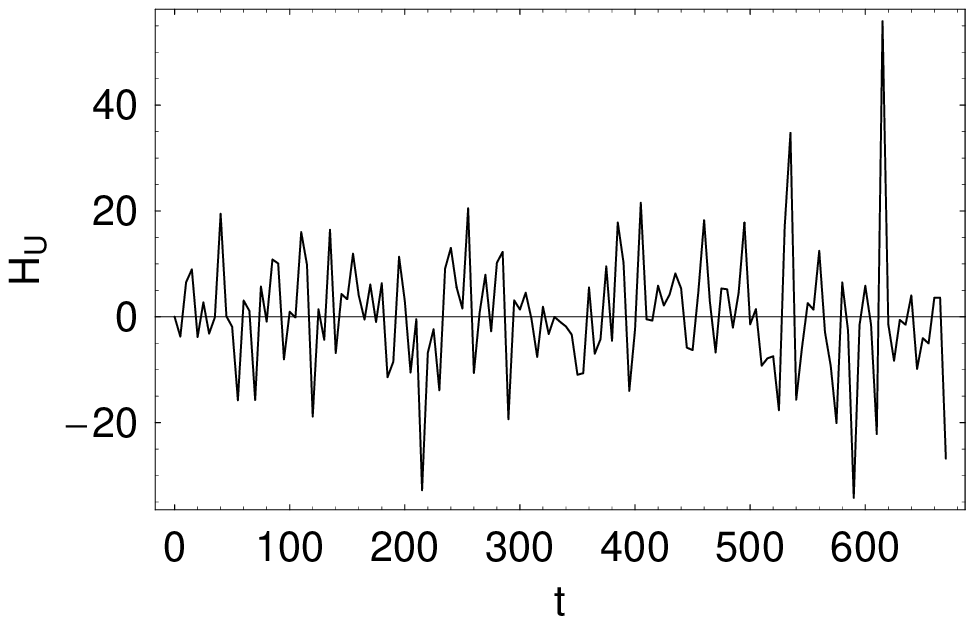}
    &
    \includegraphics[width=0.33\textwidth]{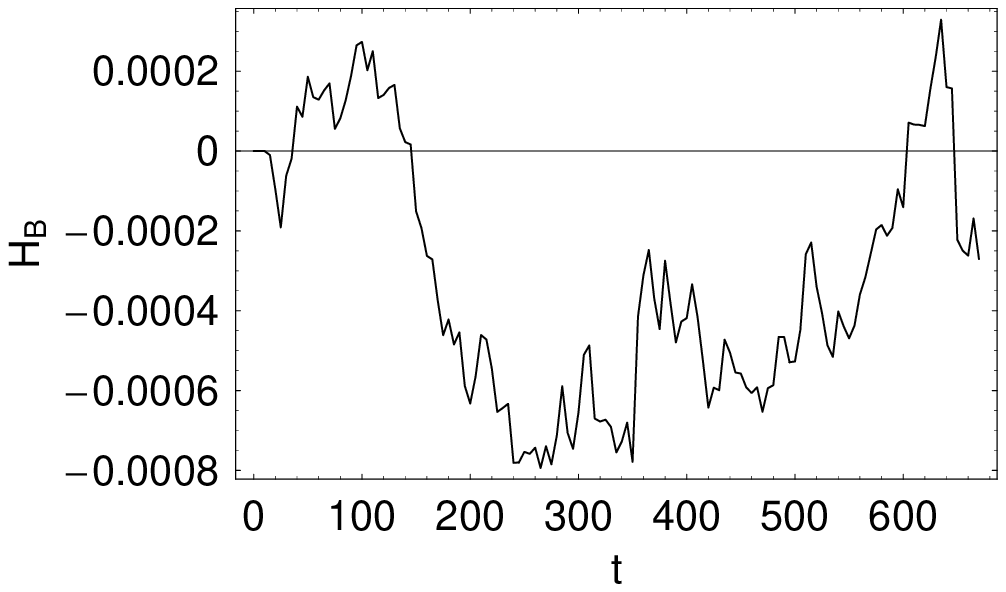}
    &
    \includegraphics[width=0.33\textwidth]{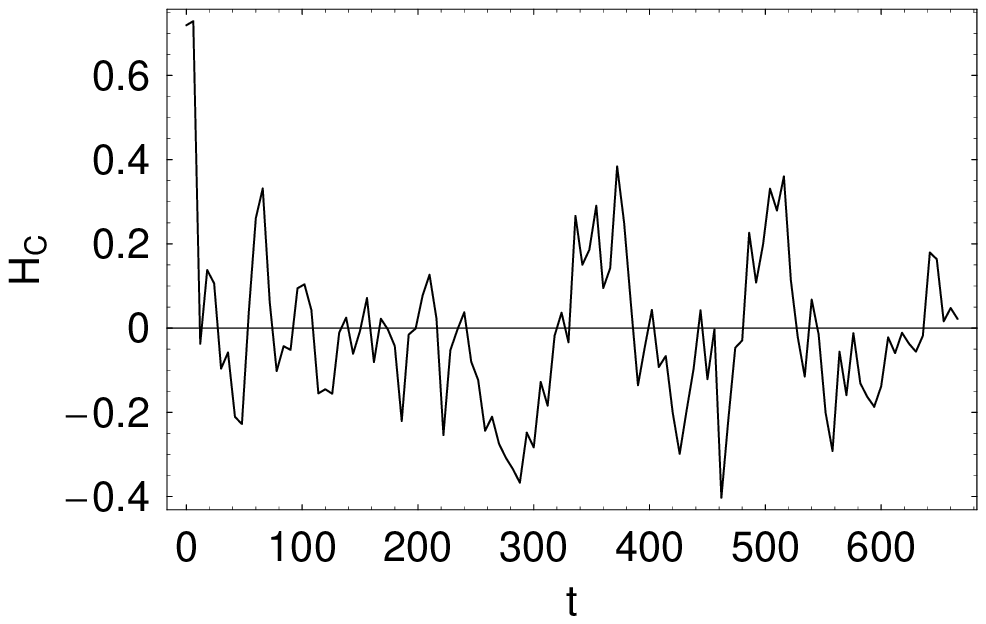}
    \\*[-0.3cm]
(d)&(e)&(f)\\*[0cm]
  \end{tabular}
\caption{Quadratic quantities (a) $E_{tot}$, (b) $E_U$, (c) $E_B$, (d) ${\cal H}_U$, (e) ${\cal H}_B$ and (f) ${\cal H}_C / \sqrt{E_U E_B}$ versus time, for $\nu=10^{-9}$ and $\eta=10^{-6}$.}
\label{figt}
\end{figure}
Here we start with a typical example of magnetic generation for
$\nu=10^{-9}$ and $\eta=10^{-6}$ ($\Pm=10^{-3}$). In Fig.~\ref{figt} the different quadratic quantities defined in
(\ref{totalE}), (\ref{crosshelicity}), (\ref{maghelicity}), (\ref{kinhelicity}) and (\ref{global}) are plotted versus time.
A coarse time sampling has been chosen here for a better representation of the results and is not relevant of the actual time step used for the numerical calculations. The kinetic, magnetic and total energies have reached a statistical stationary steady state after a few hundred time steps. The fluctuations of these quantities are quite important due to the small values of $\nu$ and $\eta$. The kinetic helicity though its average is close to zero, shows strong fluctuations.
In the other hand the magnetic helicity stays very small. Finally the relative cross helicity defined by ${\cal H}_C / \sqrt{E_U E_B}$ oscillates around zero. The fact that this latter quantity does not reach an asymptotic limit of $\pm 1$ shows that there is no ``dynamical alignement''. Therefore we are confident that our choice of forcing overcomes the problem raised by Giulani and Carbone \cite{Giulani98}.

\subsection{Spectrum analysis}\label{example}
%
\begin{figure}
  \begin{tabular}{@{}c@{\hspace{0em}}c@{}}
  \includegraphics[width=0.5\textwidth]{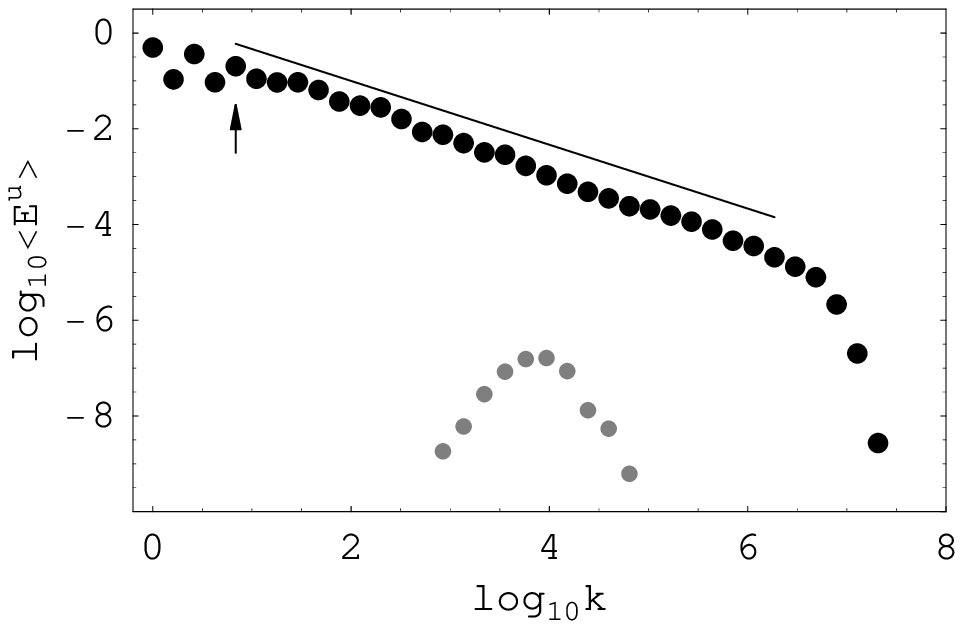}
    &
    \includegraphics[width=0.5\textwidth]{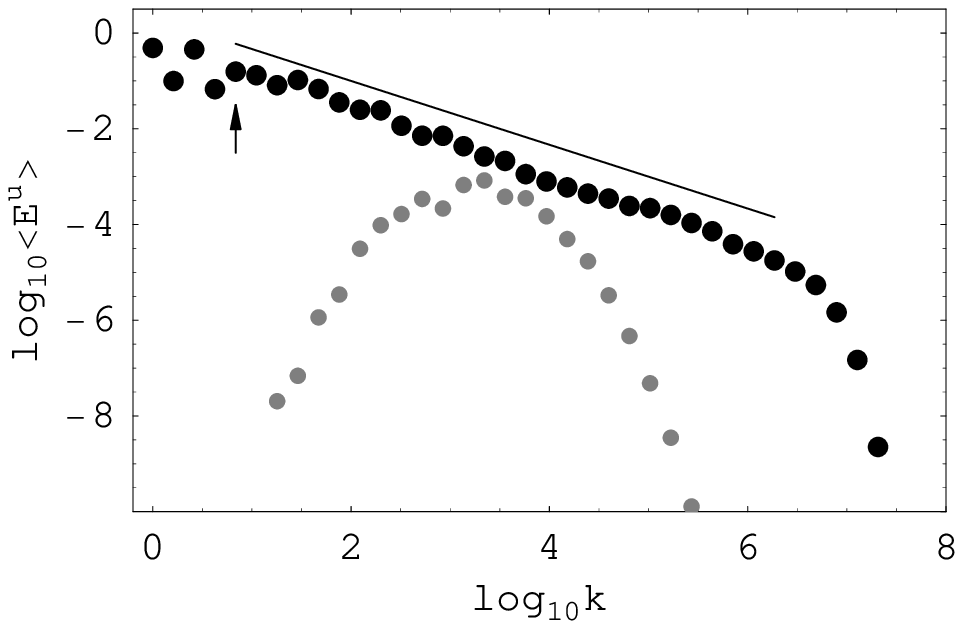}
    \\*[-2.5cm]
(a)&(b)\\*[2.5cm]
    \includegraphics[width=0.5\textwidth]{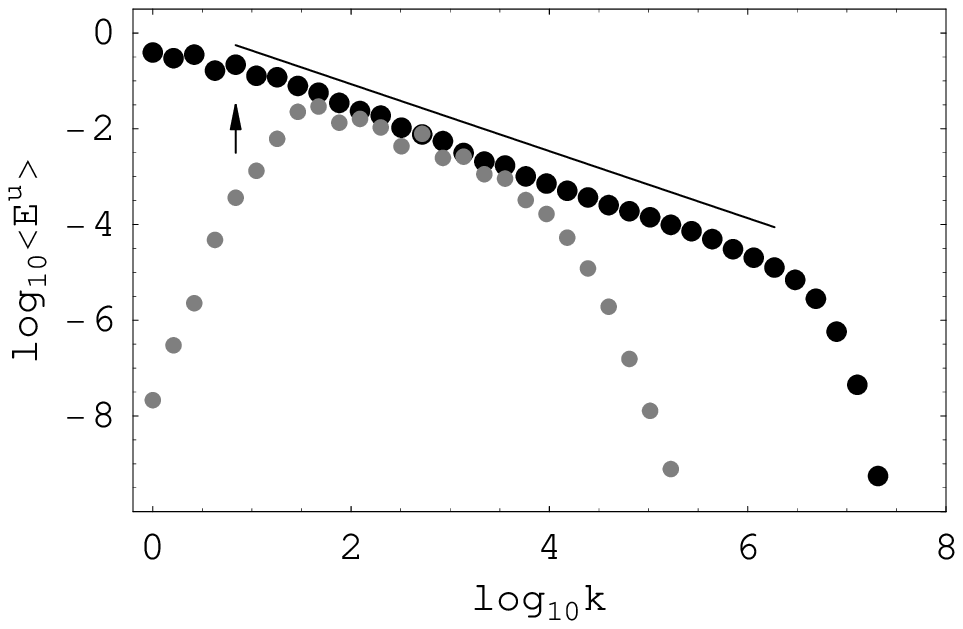}
    &
    \includegraphics[width=0.5\textwidth]{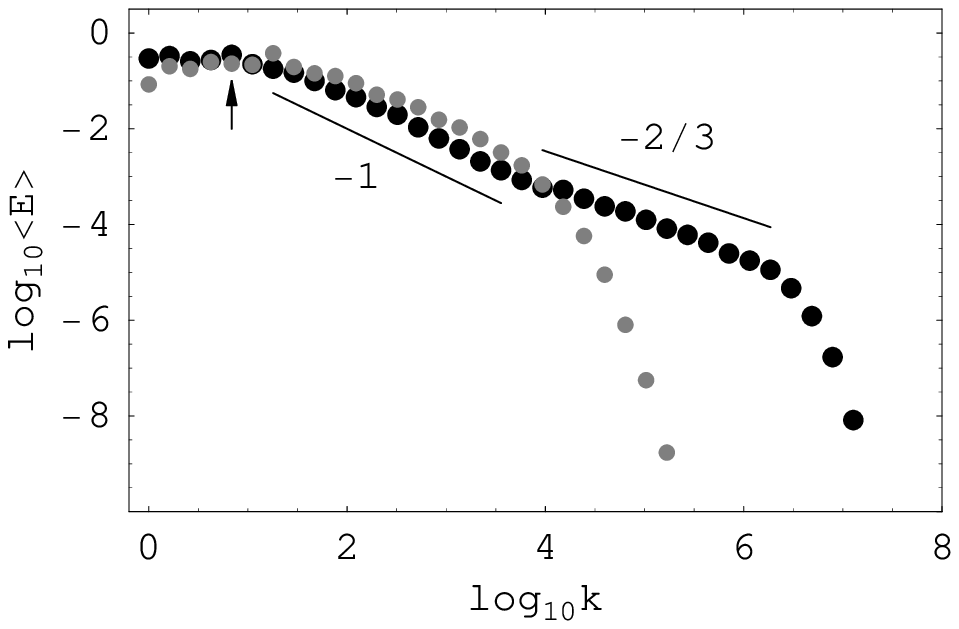}
    \\*[-2.5cm]
    (c)&(d)
    \\*[+2.5cm]
  \end{tabular}
\caption{Kinetic (black dots) and magnetic (gray dots) spectra at
four successive times (from (a) to (d)) for
$n_f=4$, $\nu=10^{-9}$ and $\eta=10^{-6}$. See also the movie energy1.mpg
in which $\log_{10} E^U(n)$ and $\log_{10} E^B(n)$ are plotted versus $\log_{10} k$ with respectively red and blue dots.} \label{fig3a}
\end{figure}
In Fig.~\ref{fig3a} we show the kinetic and magnetic spectrum at
four successive times for again $\nu=10^{-9}$ and $\eta=10^{-6}$ ($\Pm=10^{-3}$). 
Each snapshot corresponds to an average over
a not so large amount of time which explains why at early time the
kinetic spectrum is not very smooth at large scales. In the early
time, when the magnetic field is still not significant, the
kinetic energy spectrum has a slope in $k^{-2/3}$ (corresponding to a Fourier spectrum in $k^{-5/3}$).
Then, as $\Rm$ is much larger than the critical value of the
dynamo instability, the magnetic energy starts to grow (Fig.~\ref{fig3a}a). 
We expect magnetic energy to be
initially amplified by the eddies having the highest sharing rate,
i.e. the smallest scale eddies. As $\Pm < 1$, the smallest eddies
available for dynamo action correspond to eddies at resistive scale.
This is indeed what we find, as here, the resistive scale (defined as in section \ref{Inandout})
corresponds to $\log_{10}k_{\eta}\sim 4.1$.
We note that the Kolmogorov resistive scale given by
 $k_{\eta}\sim(\epsilon/\eta^3)^{1/4}$ (see section \ref{diss-scales}) with $\eta=10^{-6}$, 
leads to a slightly higher value $\log_{10}k_{\eta}\sim 4.5$. 

As $\Rm$ is sufficiently large, at subsequent times the magnetic
energy reaches the level of kinetic energy (Fig.~\ref{fig3a}c). At that time the kinetic spectrum is not
influenced yet by the nonlinear feedback of the magnetic field and
is still in $k^{-2/3}$. Then the dynamical equilibrium between the
magnetic and velocity fields settles down (Fig.~\ref{fig3a}d). A striking feature of this equilibrium is the
change of slope (from -2/3 to $\sim$ -1) of the kinetic energy spectrum
for $k\le k_{\eta}$ while the magnetic spectrum is slightly above
the kinetic spectrum. We also note that the viscous dissipation
scale has increased (the right part of the kinetic spectrum
drifting to the left). This probably comes from the fact that
 there is less energy to dissipate by viscosity than at earlier time
because of the additional Joule dissipation.

\begin{figure}
  \begin{tabular}{@{}c@{\hspace{0em}}c@{\hspace{0em}}c@{}}
    \includegraphics[width=0.32\textwidth]{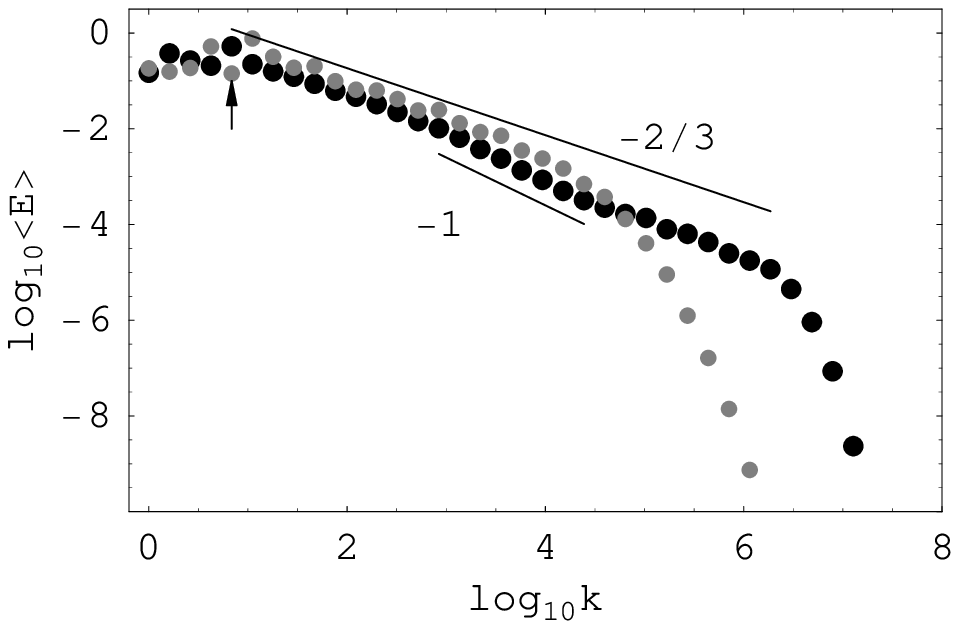}
    &
    \includegraphics[width=0.32\textwidth]{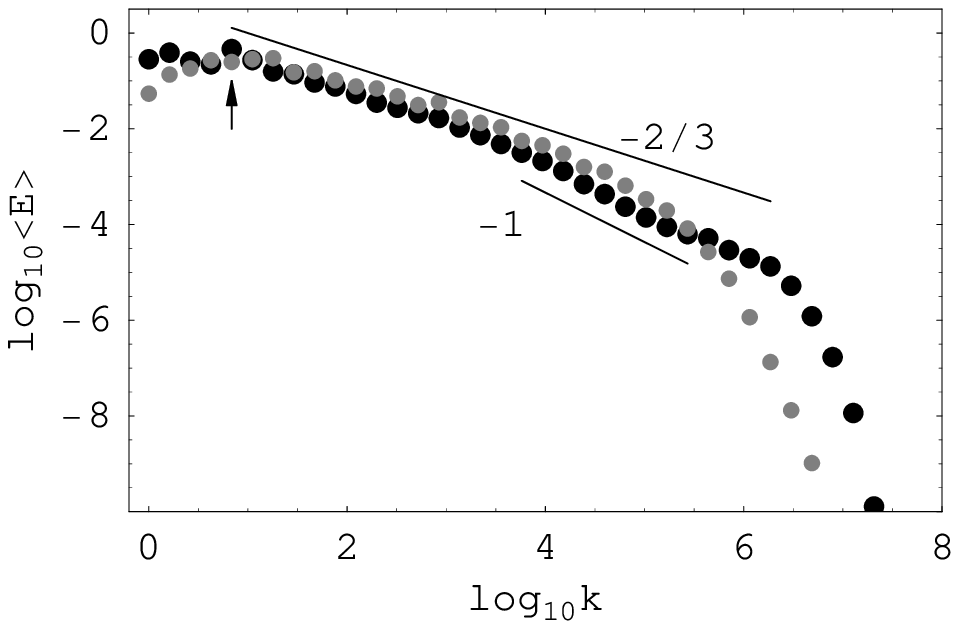}
    &
    \includegraphics[width=0.32\textwidth]{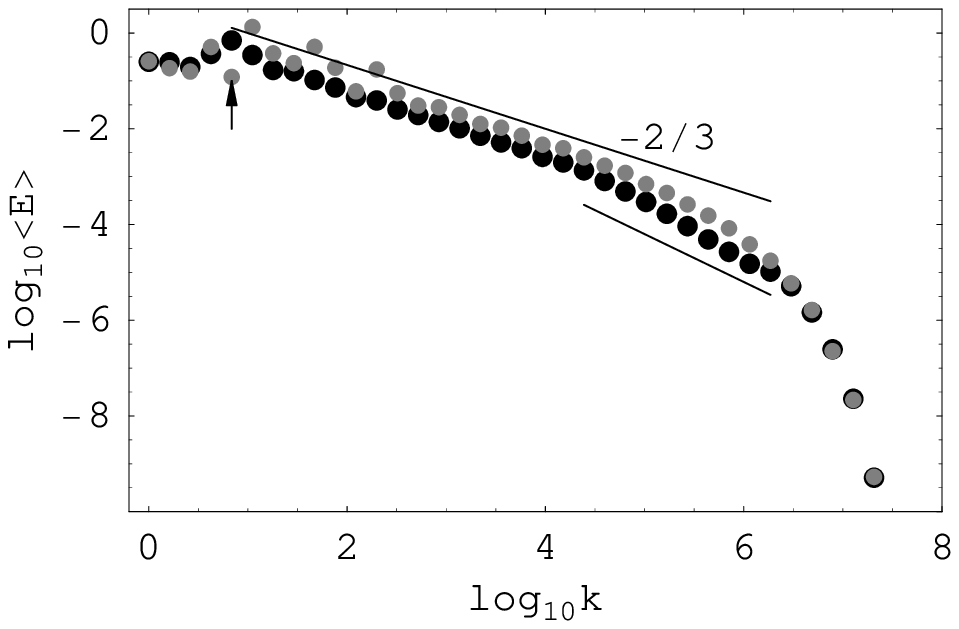}
    \\*[-1.2cm]
(a)&(b)&(c)\\*[0.5cm]
  \end{tabular}
\caption{Kinetic (black dots) and magnetic (gray dots) spectra for
$\nu=10^{-9}$ and for $\Pm=$ (a)$\;10^{-2}$,
(b)$\;10^{-1}$, (c)$\;10^{0}$ and $\Ru=$ (a)$\;6.5
\; 10^{9}$, (b)$\;4.4 \; 10^{9}$, (c)$\;4.4 \; 10^{9}$ . The forcing
scale corresponds to $n_f=4$.} \label{fig5a}
\end{figure}

\begin{figure}
      \includegraphics[width=1\textwidth]{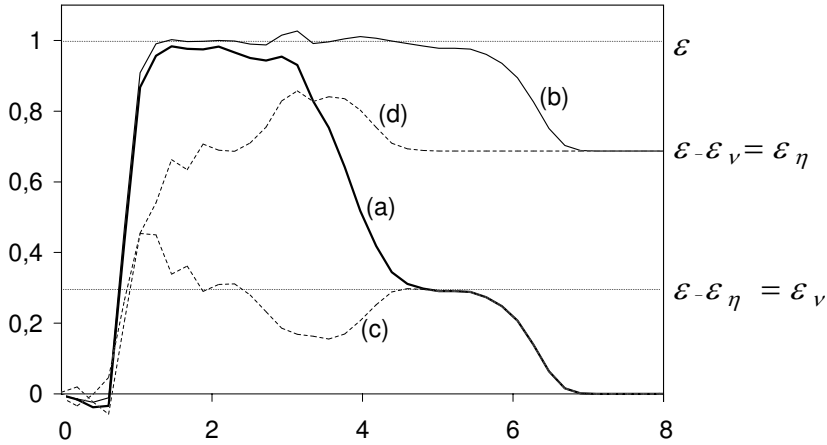}\\*[-2cm]
\caption{Spectral energy fluxes (a) $\Pi_{tot}(n)$,$\;$ (b) $\Pi_U(n)$,$\;$
 (c) $\Pi_{U>}^{U<}(n)$,$\;$
 (d) $\Pi_{U>}^{B<}(n)$
 versus $\log_{10} k$ for $\nu=10^{-9}$
and $\Pm=10^{-3}$.}
 \label{figfluxes}
\end{figure}

When changing the value of $\Pm$ while keeping the same value of $\nu$
and calculating again the final statistically stationary state, we observe
again
(Fig.~\ref{fig5a}) a deviation of the kinetic energy slope from -2/3 to $\sim$ -1 whatever the value of
$\Pm$.
To understand better these spectra, we plotted several fluxes in Fig.~\ref{figfluxes}, for $\nu=10^{-9}$
and $\Pm=10^{-3}$.

Looking at curve (a) which represents the total flux $\Pi_{tot}(n)$ versus $\log_{10} k$, one can distinguish three plateaus:
the first one corresponds to scales larger than the resistive scale
($1\le \log_{10}k \le 3$), the second one for
scales smaller than the resistive scale
but larger than the viscous scale ($\log_{10}k \sim 5$),
and the third one for scales smaller than the viscous scale
($\log_{10}k \ge 7$).
The drop from the first to the second plateau corresponds to the 
ohmic dissipation rate $\epsilon_{\eta}=\sum_{j=0}^N
D^B(j)$. The drop from the second to the third plateau corresponds to
the viscous dissipation rate $\epsilon_{\nu}=\sum_{j=0}^N D^U(j)$.
We clearly have $\epsilon = \epsilon_{\nu} + \epsilon_{\eta}$
as expected from (\ref{pitot}) for $n=N$.

The curve (b) corresponds to $\Pi_U(n)$ versus $\log_{10} k$ with two plateaus, depending if the scale is larger or smaller than the viscous scale. The first plateau ($k \le 6$) corresponds to $\Pi_U(n) \sim \epsilon$ and the second one ($k \ge 7$)
to $\Pi_U(n) \sim \epsilon- \epsilon_{\nu} = \epsilon_{\eta}$. In particular, there is no clear change of $\Pi_U(n)$ just before the resistive scale that could explain the change of slope of the kinetic energy spectrum as previously pointed out.

Now let us have a look at curve (c).
The transfer rate $\Pi_{U>}^{U<}(n)$ is responsible for the direct cascade of kinetic energy and would be constant
leading to a Kolmogorov spectrum if the magnetic field was null (see Fig.~\ref{fig1}). 
This would remain true for a non zero magnetic field only if
the curve (c) was staying flat with $\Pi_{U>}^{U<}(n)=\epsilon_{\nu}$ for
$2< \log_{10} k < 5.5$. In that case the curve (d) would be flat as well
with $\Pi_{U>}^{B<}(n)=\epsilon_{\eta}$ for $k >2$.
Instead, there is a drop of $\Pi_{U>}^{U<}(n)$ compensated by a symmetric bump of $\Pi_{U>}^{B<}(n)$
for $2< \log_{10} k < 4.5$. This drop of $\Pi_{U>}^{U<}(n)$ is consistent with a spectrum steeper than 
$k^{-2/3}$. Indeed, the bump of $\Pi_{U>}^{B<}(n)$ corresponds to some extra energy taken from $\epsilon$ and dissipated by Joule effect. Then there is less energy to be transferred through the kinetic energy cascade.
The physical reason why this scenario happens for scales just larger than the resistive scale, however is still unclear. 

For the parameters of Fig. \ref{figfluxes} the Kolmogorov
dissipation scales are given by
$k_{\eta}=(\epsilon/\eta^3)^{1/4}=10^{4.5}$ and
$k_{\eta}=(\epsilon/\nu^3)^{1/4}=10^{6.75}$ which correspond
quantitatively well with the beginning of the second and third
plateau of $\Pi_{tot}(n)$. This shows that the arguments leading
to the Kolmogorov dissipation scales (see next section) are not
affected by the change of spectra slopes observed in Fig.
\ref{fig5a}.

Finally for completeness, we produced three movies showing the time evolution of the spectra of the other quadratic quantities.
In u-helicity.mpg, b-helicity.mpg and cross-helicity.mpg,  $\log_{10} {\cal H}_U(n)$, $\log_{10} {\cal H}_B(n)$ and $\log_{10} {\cal H}_C(n)$ are plotted versus $\log_{10} k$ where the blue and red dots denote positive and negative signs.

\subsection{Dissipation scales ratio}\label{diss-scales}
\begin{figure}   \centering
\includegraphics[width=.8\textwidth]{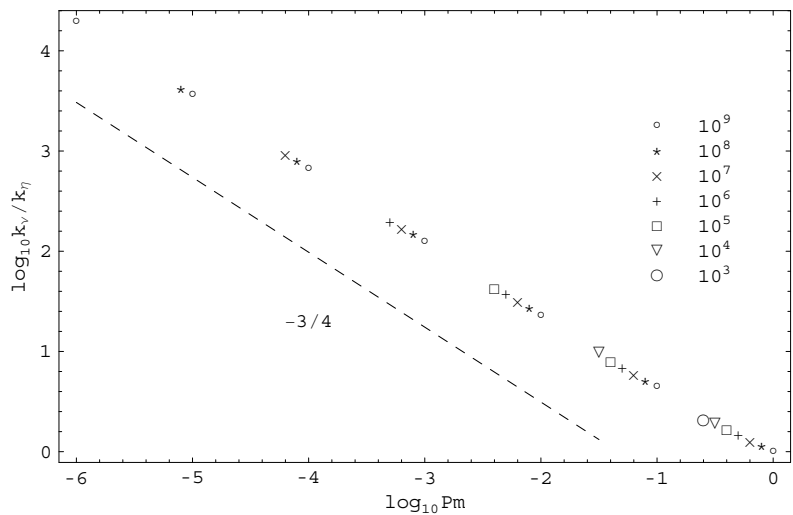}
\caption{Ratio $k_{\nu}/k_{\eta}$ versus $\Pm$ for different
values of $\nu^{-1}$ indicated in the legend. The straight line $k^{-3/4}$ is plotted (dashed line) for comparison.}\label{fig4d}
\end{figure}

At the end of section \ref{Inandout} we have already explained how we identify the 
viscous and resistive scales $k_{\nu}$ and $k_{\eta}$, by comparing the turn over time to the respective dissipative times.
In Fig.~\ref{fig4d} we plot the ratio
$k_{\nu}/k_{\eta}$ versus $\Pm \le 1$ for different values of $\Ru$.
We find that $k_{\nu}/k_{\eta} \sim \Pm^{-3/4}$. To understand why,
it is sufficient to say that between $k_{\eta}$ and $k_{\nu}$
the kinetic energy obeys a Kolmogorov spectrum $U(k)=\epsilon^{1/3} k^{-1/3}$ (see Fig. \ref{fig5a}), leading to $\tau_U^{-1}=kU(k)=\epsilon^{1/3} k^{2/3}$.
Comparing $\tau_U^{-1}$  with respectively $\tau_{\nu}^{-1}=\nu k^2$ and $\tau_{\eta}^{-1}=\eta k^2$ leads \cite{Kraichnan67} to the dissipation scales
$k_{\nu}\sim (\nu^3 / \epsilon)^{-1/4}$ and
$k_{\eta}\sim (\eta^3 / \epsilon)^{-1/4}$. This in turn leads to a 
dissipation scales ratio in $\Pm^{-3/4}$. 
\subsection{Route to saturation}
\label{saturation} 
\begin{figure}   \centering
\includegraphics[width=.8\textwidth]{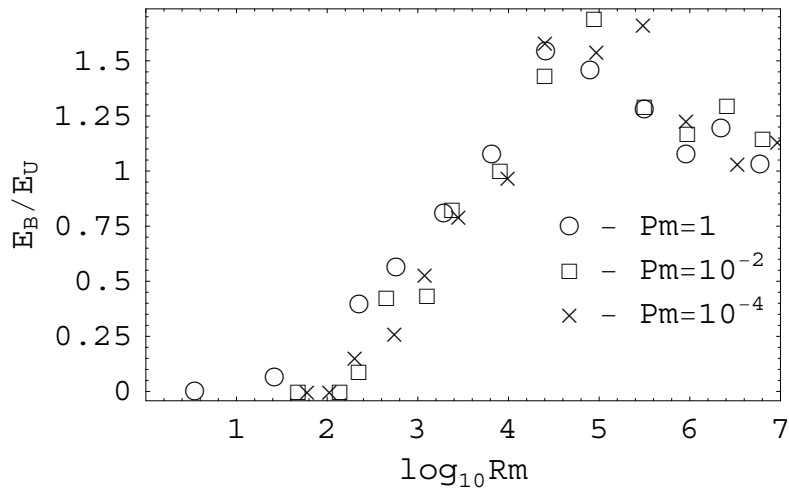}
\caption{The energy ratio $E_B / E_U$ versus $\Rm$ for $n_f=4$ and three values of
$\Pm$.}
\label{bifurcation}
\end{figure}

In this section we study the influence of $\Pm$
on the way the dynamo saturates. For that we calculate the ratio
of magnetic to kinetic energy $E_B / E_U$, $E_B$ and $E_U$ being
defined as in (\ref{global}). In Fig.~\ref{bifurcation}, $E_B /
E_U$ is plotted versus $\Rm$ for three values of $\Pm$. We note that
for $\Rm$ much larger than the critical value, the level of
saturation $E_B / E_U$ may go beyond 1 for $\Rm\sim 10^{5}$. Such a super saturation state could
be expected from the spectra of Fig. \ref{fig5a}. 
At the
threshold, the slope of $E_B / E_U$ versus $\Rm$ follows a
turbulent scaling of the form $E_B / E_U \sim (\Rm -
\Rm_c)/\Rm_c^2$ as expected by P\'etr\'elis and Fauve
\cite{Petrelis01}. Indeed as in this case the threshold $\Rm_c$
does not vary very much with $\Pm$, the slopes at $\Rm=\Rm_c$ are
similar. This is to contrast with the laminar scaling $E_B / E_U
\sim \Pm (\Rm - \Rm_c)/\Rm_c^2$ \cite{Petrelis01} which would lead
to a quasi-horizontal slope for $\Pm=10^{-4}$.
\subsection{Dynamo threshold}
\label{threshold} 
\begin{figure}   \centering
\includegraphics[width=.75\textwidth]{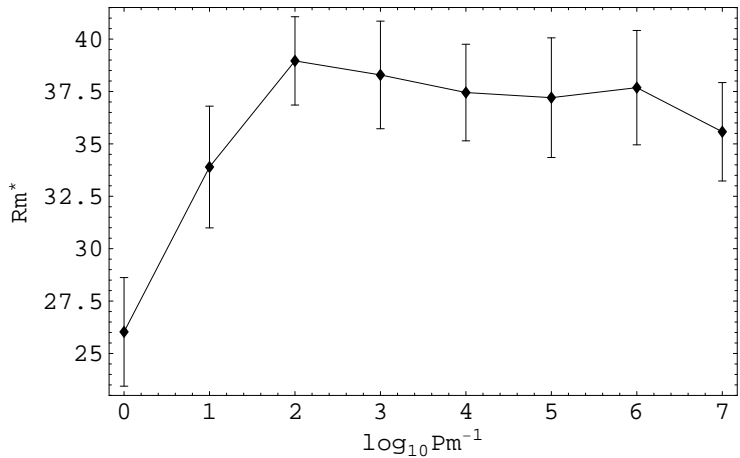}
\caption{Dynamo threshold $\Rm_c$ versus $\Pm^{-1}$ for $n_f=4$.}
\label{Rmc}
\end{figure}

In Fig.~\ref{Rmc} the dynamo threshold $\Rm_c$
is plotted versus $\Pm^{-1}$ for $n_f=4$. For increasing values of
$\Pm^{-1}$ up to $10^3$ the threshold first increases in
accordance with previous direct numerical simulations
\cite{Nordlund92,Brandenburg96b,Nore97,Christensen99,Yousef03,Scheko04}. However, for
values of
 $\Pm^{-1}$ larger than $10^3$ the threshold $\Rm_c$ is found to reach a
 plateau.
 
 For each value of $\Pm$,
 the vertical bar around $\Rm_c$ corresponds to values of $\Rm$ for which the
 magnetic solution is erratic. In other words, below the bars there is no dynamo
 action and above the bars there is a well define statistically stationary magnetic solution.
 In between though we do not observe intermittency as in \cite{Leprovost05,Leprovost06},
 the dynamo is irregular, the mean magnetic energy increasing and decreasing versus time.
\subsection{Influence of a forcing scale smaller than the resistive dissipation scale}
\begin{figure}   \centering
\includegraphics[width=.8\textwidth]{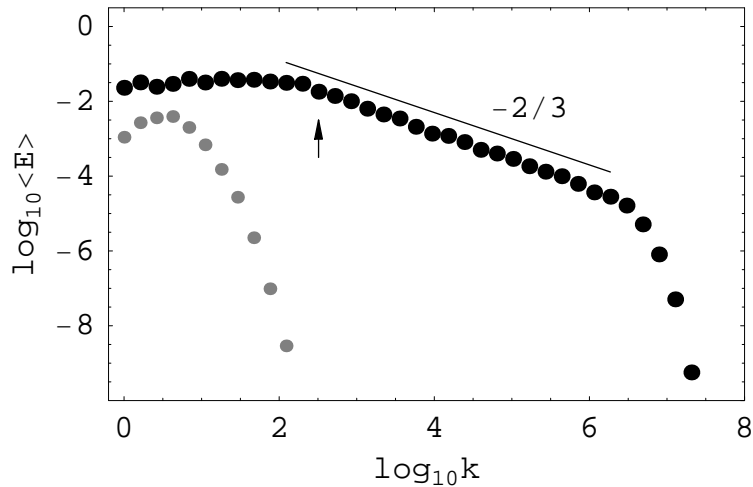}
\caption{Kinetic (black dots) and magnetic (gray dots) stationary spectra for
$\nu=10^{-9}$, $\Pm=10^{-7}$ and a forcing scale corresponding to
$n_f=12$. See also the movie energy2.mpg in which $\log_{10} E^U(n)$ and $\log_{10} E^B(n)$ are plotted versus $\log_{10} k$ with respectively red and blue dots.}\label{fig6a}
\end{figure}

In Fig.~\ref{fig6a}, the kinetic and magnetic spectra are plotted
for a forcing scale smaller than the resistive scale $k_{\eta}$. In
that case the inertial range does not play a role in the magnetic
generation and a kinetic spectra in $k^{-2/3}$ is recovered.
\section{Discussion}
In this paper we investigated the fully developed MHD turbulence at magnetic Prandtl number lower than unity,
using a shell model of MHD turbulence with an appropriate forcing.
The main results are:\\
1. For strong MHD turbulent dynamo states (large $\Rm$) we find
kinetic and magnetic energy spectra close to the Kolmogorov
spectrum $k^{-2/3}$ except at scales just larger than the
resistive dissipation scale for which there is a weaker (stronger)
slope of the kinetic (magnetic) spectrum. This corresponds to the
work of the Lorentz forces which increases with $k$ up to $k=k_{\eta}$.\\
2. The evaluation of the viscous and resistive dissipation scales
are consistent with Kolmogorov estimates
leading to $k_{\nu}/k_{\eta} \sim \Pm^{-3/4}$.\\
3. At the dynamo threshold $\Rm_c$, the ratio of magnetic to kinetic energy
scales like $E_B / E_U \sim (\Rm - \Rm_c)/\Rm_c^2$, as predicted by a turbulent scaling \cite{Petrelis01}.\\
4. At very low values of $\Pm$, the dynamo threshold $\Rm_c$ reaches a plateau.\\

Of course all these results rely on the assumption that the
interactions between the different scales of motion and magnetic
field are local interactions, each shell interacting with a few
shells above and below. We believe that this should
not make much difference as long as $\Pm$ is small, the Kolmogorov
turbulence being governed by local interaction. In the other hand
our results can not being tested against the Iroshnikov-Kraichnan 
$k^{-3/2}$ Fourier spectrum prediction \cite{Biskamp03} resulting from non local
interactions between the flow and some large scale magnetic field
which could result for example from dynamo action. By the way we
believe that the $k^{-3/2}$ slope in FS98
is due to a lack of statistics as can be seen from
the energy fluxes which are not flat and from the 
corresponding small range of scales. Adding some non local interaction
with a large scale magnetic field in a local shell model,
 Biskamp \cite{Biskamp94} found a $k^{-3/2}$ slope, though taking only one such a non local interaction
 is somewhat artificial. Recently Verma \cite{Verma99} revisited the Iroshnikov-Kraichnan
 theory in which he shows that the large scale magnetic field becomes renormalized due
 to the nonlinear term, leading back to the Kolmogorov spectrum. This emphasizes
 the need for a complete nonlocal shell model in which any shell could interact with the others.
 This could be a good test against one theory or the other.
 Such a model would be also welcome for simulations at large $\Pm$.
 Indeed at large $\Pm$ we expect the more energetic scales of the flow,
 corresponding to scales close to the viscous scales, to interact directly with the
 smaller scales of the magnetic field. Our local shell model can not catch such features
 and this is why we did not show results at large $\Pm$ for they surely lack physical ground.
 A further issue that could be addressed
 by a nonlocal shell model could be to distinguish between a large scale field generated
 by a small scale velocity field resulting from non local interactions
 (developed in the mean field formalism) and a large scale field generated
 by an ''inverse cascade'' as for example in Fig. \ref{fig3a} or in \cite{Pouquet76},
 resulting from local interactions.\\
Concerning our local model, we believe that the results presented in Fig. \ref{Rmc}
showing that the dynamo threshold does not depend on $\Pm$ at low
values of $\Pm$ would stay qualitatively the same if additional
nonlocal interactions were included in the model. Indeed the
dynamo threshold corresponds to the growth start of the magnetic
field which is then still not significant. Therefore any non local
interactions (e.g. Alfven sweeping effect) might not change the
threshold.
\\

\section{Acknowledgments}
Most of this work was done during a stay of R. S. at the LEGI,
with a grant from the Universit\'e Joseph Fourier, Grenoble,
France and completed during the visit of F.P. at the ICMM, Perm,
Russia, supported by the ECO-NET program 10257QL.
R.S. is also thankful for support from the BRHE program.
\section{Appendix}
For the pure hydrodynamic case ($B_n=0$), only the two first conditions (\ref{forcing1}) and (\ref{forcing2})
are necessary to derive the forcing equations. In that case the forcing set writes
\begin{eqnarray}
    f_0 &=& \frac{\lambda \varepsilon}{(\lambda+1)u_0\cos(\phi_0-\omega_0)} \label{f1}\\
    f_1 &=& \frac{\varepsilon}{(\lambda+1)u_1\cos(\phi_1-\omega_1)} \label{f2}\\
    f_2 &=& 0, \label{f3}
\end{eqnarray}
while for the full MHD case the forcing set is derived from the
three conditions (\ref{forcing1}), (\ref{forcing2}) and
(\ref{forcing3})
\begin{eqnarray}
\frac{A}{\varepsilon}( 1 + \lambda)f_0 &=& \lambda b_2 u_1 \cos (\theta_2 - \phi_2)\,\cos (\phi_1 - \omega_1)\nonumber\\
                    &+& \lambda^2 b_1 u_2 \cos (\theta_1 - \phi_1)\,\cos (\phi_2 - \omega_2) \label{f1MHD}\\
\frac{A}{\varepsilon}( 1 + \lambda)f_1 &=& b_2 u_0 \cos(\theta_2-\phi_2)\cos(\phi_0-\omega_0)\nonumber\\
                    &-& \lambda^2 b_0 u_2 \cos(\theta_0-\phi_0)\cos(\phi_2-\omega_2) \label{f2MHD}\\
\frac{A}{\varepsilon}( 1 + \lambda)f_2 &=& - b_1 u_0 \cos(\theta_1-\phi_1)\cos(\phi_0-\omega_0)\nonumber\\
                    &-& \lambda b_0 u_1 \cos(\theta_0-\phi_0)\cos(\phi_1-\omega_1) \label{f3MHD}
\end{eqnarray}
where
\begin{eqnarray}
    &A& = b_2 u_0 u_1 \cos(\theta_2 - \phi_2) \cos(\phi _0 - \omega_0) \cos(\phi_1 - \omega_1)\nonumber\\
      &+& (\lambda -1) b_1 u_0 u_2 \cos(\theta_1 - \phi_1) \cos(\phi _0 - \omega_0) \cos(\phi_2 - \omega_2) \nonumber\\
      &-& \lambda b_0 u_1 u_2 \cos(\theta_0 - \phi_0) \cos(\phi _1 - \omega_1) \cos(\phi_2 - \omega_2)
\end{eqnarray}
and where $u_j$ and $\omega_j$ (resp. $b_j$ and $\theta_j$) are the complex modulus and argument of $U_{n_f+j}$ (resp. $B_{n_f+j}$).

\end{document}